\documentstyle[12pt,cite,epsfig,psfrag]{article}

\def\be{\begin{equation}}
\def\ee{\end{equation}}
\def\bea{\begin{eqnarray}}
\def\eea{\end{eqnarray}}

\newcommand{\beqal}{\begin{eqnarray}\label}
\newcommand{\beqa}{\begin{eqnarray}}
\newcommand{\eeqa}{\end{eqnarray}}

\begin{document}
\baselineskip=0.6cm
\begin{titlepage}
\begin{center}
%\hfill hep-th/yymmnnn\\
%\hfill IP/BBSR/2008-??\\
\vskip .2in

{\Large \bf Bouncing Braneworld with Born-Infeld and Gauss-Bonnet}
\vskip .5in

{\bf Sudipta Mukherji$^*$\footnote{e-mail: mukherji@iopb.res.in} and
Supratik Pal$^+$\footnote{e-mail: supratik\_v@isical.ac.in}\\
\vskip .1in
{\em $^*$Institute of Physics,\\
Bhubaneswar 751 005, India.}
\vskip .1in
{\em $^+$Physics and Applied Mathematics Unit,\\
Indian Statistical Institute,\\
203 B.T. Road, Kolkata 700 108, India.}}

\end{center}

\begin{center} {\bf ABSTRACT}
\end{center}
\begin{quotation}

\baselineskip 15pt

\noindent We show the existence of some bouncing cosmological solutions in the 
braneworld 
scenario. More specifically, we consider a dynamical three-brane in the background 
of Born-Infeld and electrically charged Gauss-Bonnet black hole.
We find that, in certain range of parameter space, the brane universe, at least 
classically, never shrinks to a zero size, resulting in  a singularity-free  cosmology 
within the classical domain.
\end{quotation}
\vskip 2in
16 June 2009\\
\end{titlepage}
\vfill
\eject

Singularity-free cosmological solutions are very rare within the 
framework of the conventional Einstein 
theory of gravity. It is however believed that at a short distance scale, Einstein 
action would require modifications. These modifications can, in principle, have 
cosmological consequences and may induce nonsingular cosmologies by evading the 
singularity theorem \cite{Hawking:1969sw}. Indeed, in string theory, one expects 
drastic modifications of Einstein action at high energies. 
These modifications appear due to worldsheet corrections and 
due to contributions from the quantum loops. These corrections  may violate 
strong energy 
condition, making it possible to construct non-singular time dependent 
solutions \cite{Antoniadis:1993jc, Gasperini:2002bn, Kachru:2002kx}. 

Physics at a  small scale  receives modifications in the braneworld 
scenario as well. 
In the cosmological sector, this is substantiated via a modified version
of the Friedmann equations resulting from the embedding geometry.
Indeed in many braneworld models, it has been found that the universe can have 
a smooth transition from a contracting to an expanding phase without reaching 
a singularity \cite{Mukherji:2002ft, Biswas:2003rb, Shtanov:2002mb, 
DeRisi:2007dn}. 
In particular, in \cite{Mukherji:2002ft}, a non-singular cosmological 
solution was 
presented by considering a dynamical three-brane in a five dimensional charged AdS 
black hole background. This happens because the charge of the bulk black hole induces a 
negative energy 
density on the brane world volume. This, in turn, prevents the brane to fall in to 
the black hole singularity. However, this bounce, unfortunately, occurs inside the 
Cauchy horizon of the black hole. Consequently, regardless of the initial
energy of the brane, the bulk space-time collapses due to the back-reaction at the 
Cauchy horizon \cite{Hovdebo:2003ug}.

Within the framework of braneworld scenario, in this short note, we search for 
other probable models, where, at least classically, one can construct non-singular 
cosmological
universe. In the following, we show the existence of such solutions by 
considering an
empty three-brane in the background of 
Born-Infeld black hole \cite{Dey:2004yt} and in the 
background 
of charged Gauss-Bonnet (GB) black hole \cite{Lidsey:2002zw, Torii:2005nh, Dey:2007vt}. In both the cases, by 
considering the 
effective potentials, we show that, within a certain range of parameter space,  
there do exist 
bouncing solutions in these backgrounds as well. 
An important conclusion of this analysis is that the charge of the bulk black hole plays the key role
in avoiding cosmological singularity on the brane.
It is however important to study the stability of such 
solution. We have not done this in this note and expect to come back to it 
in the future.

\bigskip

\bigskip

\noindent {\bf {Bounce with Born-Infeld}}

\bigskip

\bigskip

The Einstein-Born-Infeld action in five dimensions in the presence of a 
negative bulk cosmological constant $\Lambda$ is given by
\begin{equation}
S = \int d^5x\sqrt{-g}\Big[{1\over{\kappa}}(R - 2 \Lambda) + L(F)\Big],
\end{equation}
where $L(F)$ is the contribution from the Born-Infeld gravity
\begin{equation}
L(F) = 4 \beta^2 \Big(1 - {\sqrt{1 + {F_{\mu\nu} 
F^{\mu\nu}\over{2\beta^2}}}}\Big).
\end{equation}
Here the constant $\beta$ is the Born-Infeld parameter, having the dimension of 
mass. In the limit $\beta \rightarrow \infty$, $L(F)$ reduces to the 
standard Maxwell form 
\begin{equation}
L(F) = - F^{\mu\nu} F_{\mu\nu} + {\cal {O}} (F^4).
\end{equation}
A $(n+1)$ dimensional black hole solution of this theory has  been obtained 
in \cite{Dey:2004yt} by considering the 
general form of the metric as 
\begin{equation}
ds^2 = - V(r) dt^2 + {dr^2\over{V(r)}} + r^2 d\Omega^2_3,
\label{bbb}
\end{equation}
which gives,  after some rigorous calculations, a solution for the metric function 
$V(r)$ in $(n+1)$ dimensions \cite{Dey:2004yt}.
In $(4+1)$ dimensions, the metric function $V(r)$ can be written in a compact form in terms of Hypergeometric  
and algebraic functions as
\begin{eqnarray}
V(r) = 1 - {m\over r^2} + \Big[{\beta^2\over 3} + {1\over l^2}\Big] r^2
    -{ {\sqrt 2}\beta \over {6 r}} {\sqrt{2 \beta^2 r^{6} + 6 
q^2}}
    + {3 \over{2 r^4}} {q^2} ~ {}_2F_1\big[{1\over 3}, {1\over 2}, {4\over 3}, 
      - { 3 q^2\over{\beta^2 r^{6}}}\Big].
\label{biv}
\end{eqnarray}
The solution of the corresponding gauge field equations of motion is
\begin{equation}
F^{rt} = { \sqrt{3} \beta q\over{\sqrt{\beta^2 r^6 + 3 q^2}}},
\label{bif}
\end{equation}
with all the other components of $F^{\mu\nu}$ being zero.
In (\ref{biv}), $m$ and $q$ are related to the ADM mass and charge of the black hole respectively. The 
constant $l$, appearing in various equations, parametrises the cosmological 
constant $\Lambda$ 
as $\Lambda = - 6/l^2$. The asymptotic behaviour (large $r$ limit) of the metric function
can be obtained by using the properties of the Hypergeometric function in (\ref{biv}), which 
gives
\begin{equation}
V(r)|_{r \rightarrow \infty} = 1 - {m\over r^2} + {q^2\over{r^4}} + {r^2\over l^2} - { 3 q^4\over{ 16 \beta^2 
r^{10}}}.
\label{asymp} 
\end{equation}
Note that in the limit of large $\beta$, it has the form of Reissner-Nordstrom AdS black hole,
which is obvious from the fact that, in the large $\beta$ limit, the non-zero gauge field component
reduces to
\begin{equation}
 F^{r t} |_{\beta \rightarrow \infty} \sim \frac{q}{r^3}.
\end{equation}

It is instructive to consider the nature of the horizon(s) 
associated with this black hole. This is perhaps best described 
by a plot of the mass of the hole as a function of the  
horizon radius. The horizons correspond to the zeros of $V(r)$. Denoting the radius by $r_h$,
from (\ref{biv}), we get
\begin{equation}
m = r_h^2 + \Big[{\beta^2\over 3} + {1\over l^2}\Big] r_h^4
    -{ {\sqrt 2}\beta r_h\over {6}} {\sqrt{2 \beta^2 r_h^{6} + 6
q^2}}
    + {3 \over{2 r_h^2}} {q^2} ~ {}_2F_1\big[{1\over 3}, {1\over 2}, {4\over 3},
      - { 3 q^2\over{\beta^2 r_h^{6}}}\Big].
\label{mrb}
\end{equation}

\begin{figure}[t]
\begin{center}
\begin{psfrags}
\psfrag{a}[][]{$ r_h$}
\psfrag{b}[][]{$m$}
\epsfig{file=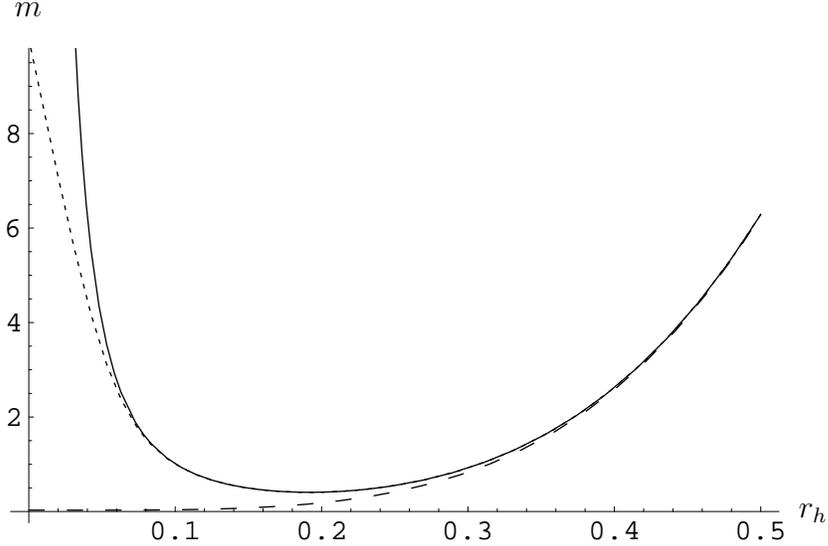, width= 12cm,angle=0}
\end{psfrags}
\vspace{ .1 in }
\caption{Plot for the mass of BI black hole as a function of horizon radius for fixed charge.
The dashed 
line is for $\beta < \beta_c$ and the rest two are for $\beta > \beta_c$. 
For a specific choice of parameters $q = 0.1$ and $l = 1$, the curves correspond to $\beta = 0.1$ (dashed),
635 (dotted) and $10^{5}$ (continuous line) respectively.}
\label{bimb}
\end{center}
\end{figure}

In figure (\ref{bimb}), we have plotted $m(r_h)$ for fixed $q$ and for different values 
of $\beta$. For a given mass, we clearly see that the number of horizons depend on the
value of $\beta$. For fixed $q$, there is a single horizon unless $\beta$ is above a 
critical value, which we call $\beta_c$. This critical value $\beta_c$ 
can be found out as follows. From the figure, we see that if we decrease the mass, 
at some value of $m$, two horizons meet. We call this an extremal hole\footnote{Black 
hole then has zero temperature.}. This minimum can be found by 
extremising
(\ref{mrb}). At the extremal point, $r_h$ then satisfies
\begin{equation}
2 r_h + 4 ({\beta^2 \over 3} + {1\over l^2})r_h^3 - {2 \sqrt 2\over 3} \beta
{\sqrt{2 \beta^2 r_h^{6} + 6 q^2}}  = 0.
\end{equation}
It can be checked that, for a given $q$, there exists a real solution for $r_h$ 
only
when $\beta$ is larger than a critical value. We previously called 
this $\beta_c$.
It can also be checked that $\beta_c$ decreases as we increase $q$,
which reveals the important role of the charge parameter in the bouncing
cosmology in the braneworld context.

We would now like to study cosmology of an empty three-brane moving in this
background. In what follows, we will show the existence of a bouncing cosmology on the brane for
$\beta > \beta_c$.

Let us start with a three-brane with tension $\Lambda_{\rm br}$ moving in the geometry
given by (\ref{bbb}) with $V(r)$ given in (\ref{biv}). The brane metric, consistent with
the orthonormality conditions and junction conditions for embedding,
is compatible to FRW geometry 
\begin{equation}
ds_4^2 = - d\tau^2 + a(\tau)^2 d\Omega_3^2,
\label{frw}
\end{equation}
$\tau$ being the proper time on the brane. Using the junction conditions relating
the extrinsic curvature with the bulk quantities,
one can obtain an FRW metric on the brane of the above form if one identifies the scale factor $a(\tau)$
on the brane  
with the radial trajectory $r(\tau)$ of the bulk black hole. From now on we shall use $a$ for $r$,
wherever it appears.
The embedding geometry thus gives \cite{Barcelo:2000re}
\begin{equation}
{\bar\Lambda_{\rm br}\over 3} = \bar H^2 + {V(\bar a)\over{\bar a^2}}. 
\label{hubBI} 
\end{equation}
%Here $\bar \Lambda_{\rm br}$ is the effective brane cosmological
%constant, which is the combined effect
%of the bulk cosmological constant $\Lambda$ and the brane tension.
In the above, 
the quantities with bar are all dimensionless and are defined as
\begin{equation}
\bar \Lambda_{\rm br} = l^2 \Lambda_{\rm br}, ~ \bar a = {a\over l}, ~\bar H = {d\bar 
a\over{\bar a d\bar \tau}}.
\end{equation}

In the limit $\beta \rightarrow \infty$, (\ref{hubBI}) reduces to
\begin{equation}
{\bar \Lambda_{\rm br}\over 3} = 1 + \bar H^2 + {1\over {\bar a^2}} - {\bar m\over
\bar a^4} + {{\bar q^2}\over {\bar a^6}}.
\label{smal}
\end{equation}
Here $\bar m = m/l^2, \bar q = q/l^2$.
This equation has a simple solution when we tune the brane tension such that
$\bar \Lambda_{\rm br} = 3$. This corresponds to setting the brane tension same as
that of the curvature radius of the asymptotic AdS space. In conformal gauge, the
solution of (\ref{smal}) is
\begin{equation}
\bar a(\bar \eta) = {\sqrt{\bar m\over 2}}\Big[1 - {\sqrt{1 - {4 \bar q^2\over {\bar
m^2}}}} {\rm cos} (2 \bar \eta)\Big],
\label{betainfty} 
\end{equation}
where $\eta$ is the conformal time defined as $d\tau = a(\eta) d\eta$ and $\bar \eta
= \eta/l$. For $\bar m > 2 \bar q$, this generates a non-singular cyclic cosmology with
brane universe reaching a minimum radius at $\eta = n \pi$ for $n$ being integer,
while the  maximum radius is reached at $\eta = (n+1/2) \pi$. 

For finite $\beta$, it is difficult to find an exact closed form solution of (\ref{hubBI}), which now includes  terms involving $\beta$ in the Friedmann equation and, in turn, make the results more complicated.
It is obvious from  the inclusion of the Hypergeometric function which does not have a
simple algebraic expression for a general value of $\beta$ as such. 
However, taking the asymptotic behaviour (\ref{asymp}) and rearranging terms, we can arrive at the following expression
for the modified Friedmann equations on the brane for large $\bar a$ limit :
\begin{equation}
{\bar \Lambda_{\rm br}\over 3} = 1 + \bar H^2 + {1\over {\bar a^2}} - {\bar m\over
\bar a^4} + {{\bar q^2}\over {\bar a^6}} - \frac{3\bar q^4}{16 \beta^2 {\bar a}^{12}},
\label{frwBI}
\end{equation}
where the term involving $\bar m$  is the usual dark radiation, the one with ${\bar q}^2$ 
gives a geometric stiff matter-like contribution (obtained in the case of infinite $\beta$ as well)
and the one with ${\bar q}^4$ and $\beta$ gives another geometric contribution (stiffer matter), which reflects the precise role
of a finite Born-Infeld parameter in this context. 
We have been able to obtain an analytical solution for the scale factor
at least in this asymptotic limit but the expressions are too lengthy to produce here.
As apparent, complicated equations sometime defer us from obtaining a simple 
analytical result but it is more important to overcome the hindrance and to extract  the 
underlying physics than to obtain a simple  analytical solution. 

So, irrespective of whether or not there is  a simple analytical solution for a general
Born-Infeld parameter, 
it is rather more relevant to ask whether we can have a
 bouncing cosmology in this Born-Infeld background with a finite $\beta$ as well.
In the following, we incorporate a mechanism by analysing the
 effective potential  to find that
the   bouncing solution still survives if $\beta 
> \beta_c$.
In order to see this, we rewrite (\ref{hubBI}) as 
\begin{equation}
\Big({d\bar a\over{d\bar \tau}}\Big)^2 + U(\bar a) = 0,
\end{equation}
Here $U(\bar a)$ is the effective potential given by
\begin{equation}
 U(\bar a) = V(\bar a) - {\bar a}^2.
\end{equation}
where we have set, as before, $\bar\Lambda_{\rm br} = 3$.

\begin{figure}[t]
\begin{center}
\begin{psfrags}
\psfrag{a}[][]{$\bar a$}
\psfrag{b}[][]{$U(\bar a)$}
\epsfig{file=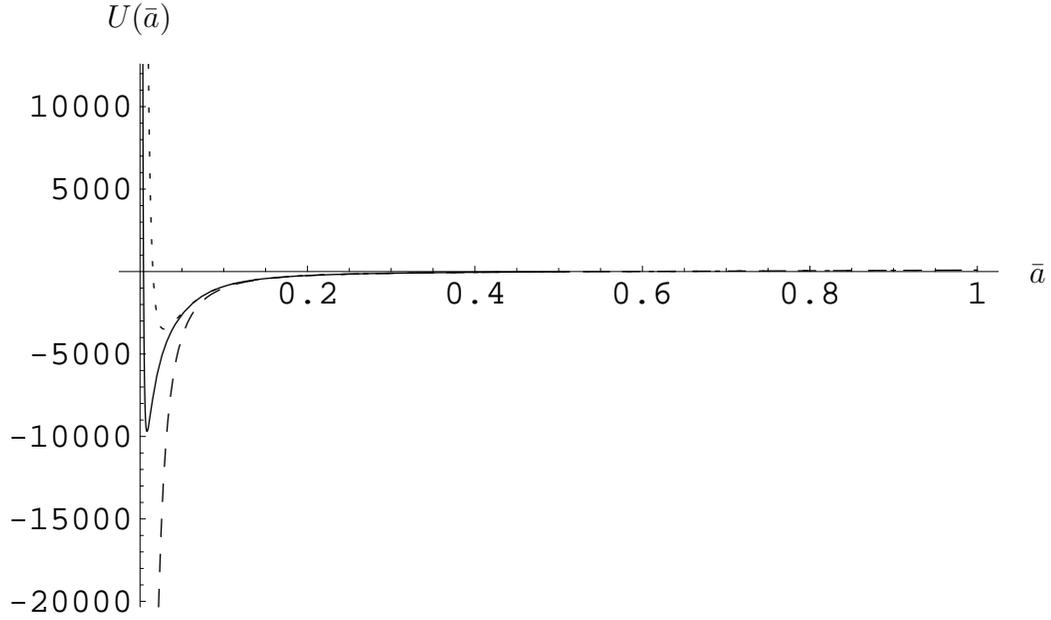, width= 14cm,angle=0}
\end{psfrags}
\vspace{ .1 in }
\caption{Plot of effective potential $U(\bar a)$ as a function of $\bar a$. The dashed line is for 
$\beta < \beta_c$ and the rest two are for $\beta > \beta_c$. In this plot, for a specific set of
$\bar m = 10$, $\bar q = 0.1$ we have $\beta = 0.1$ (dashed), 635 (continuous line) and 
$1000$ (dotted) respectively, with $\beta = 635$ being its critical value
for this choice of parameters.}
\label{ubi}
\end{center}
\end{figure}

For $\beta < \beta_c$ and $\beta > \beta_c$, the plots are 
shown in figure (\ref{ubi}). Physical domain of the solution corresponds to the 
region $U(\bar a) \le 0$ with $U(\bar a) = 0$ representing the turning points.
From the plots we see that for very small $\beta$, the effective potential does not
show any signature for bounce. On the contrary, above a specific value for the
BI parameter $\beta = \beta_c$, there are, indeed, bouncing cosmological solutions
as indicated by the nature of the
effective potential. We have previously termed this $\beta_c$ as the critical value
for the BI parameter. It readily follows from the above analysis that,
 so far as the BI parameter is above a critical
value, we have non-singular cosmology on the brane. This is true for infinite value
of $\beta$ as well, as supported by the analytical solution  (\ref{betainfty}).

In the rest of this note, we show that even in the Gauss-Bonnet theory, in the 
presence of electric charge, it is possible to construct a bouncing cosmological 
solution.

\bigskip

\bigskip

\noindent{\bf{Bounce with Gauss-Bonnet}}

\bigskip

\bigskip

The Gauss-Bonnet action with a Maxwell term is given by
\begin{equation}
S = \int d^5x\sqrt{-g}\Big[{R\over \kappa} - 2 \Lambda + \alpha (R^2 - 4 R_{ab}R^{ab}
+ R_{abcd}R^{abcd} ) - {F^2\over \kappa}
 \Big].
\label{act}
\end{equation}
Here, $\alpha$ is the GB coupling. This action possesses black hole solutions carrying 
electric charge \cite{Torii:2005nh, Dey:2007vt}. These solutions have the form
\begin{equation}
ds^2 = -V(r) dt^2 + {dr^2\over{V(r)}} + r^2 d\Omega^2_3,
\label{bh}
\end{equation}
where $V(r)$ is given by
\begin{equation}
V(r) = 1 + {r^2\over{2 \hat\alpha}} - {r^2\over{2 \hat\alpha}}\Big[1 - {4 \hat 
\alpha\over{l^2}}
+ {4 \hat\alpha m\over {r^4}} - {4 \hat\alpha q^2\over{r^6}}\Big]^{1\over 2}.
\label{vblack}
\end{equation}
In the above expression, $\hat\alpha = 2 \alpha \kappa$, while $l$ is related to the 
cosmological constant as $l^2 = - 6/(\kappa \Lambda)$. As before, the parameters $m$ and $q$ are 
related to the ADM mass and the charge respectively. The gauge potential is
given by
\begin{equation}
A_t = - {\sqrt{3}\over 2} {q\over r^2} + \Phi,
\end{equation}
where $\Phi$ is a constant which we will fix below. Denoting $r_+$ as the largest real 
positive root of $V(r)$, we find that the metric (\ref{bh}) describes a black hole with 
a non-singular horizon if
\begin{equation}
2 r_+^6 + l^2 r_+^4 \ge q^2 l^2.
\end{equation}
In the following we would choose the gauge potential $A_t$ to vanish at the horizon, 
which, in turn, fixes $\Phi$ as 
\begin{equation}
\Phi = - {\sqrt{3}\over 2} {q\over r_+^2}.
\end{equation}
Asymptotically, the metric (\ref{bh}) is an AdS space with
\begin{equation}
V(r)|_{r\rightarrow \infty} = 1 + \Big[ {1\over {2 \hat\alpha}} - {1\over {2 \hat\alpha}}
\Big(1 - { 4 \hat\alpha\over{l^2}}\Big)^{1\over 2}\Big] r^2.
\end{equation}
Therefore, the effective AdS length associated with the asymptotic metric
is given by 
\begin{equation}
L^2 = \Big[ {1\over{2 \hat\alpha}}- {1\over {2\hat\alpha}}(1 - { 4 
\hat\alpha\over{l^2}}\Big)^{1\over 2}\Big]^{-1}
\label{efads}
\end{equation} 
We now note that the metric is real if,
\begin{equation}
\hat\alpha \le {l^2\over 4}.
\end{equation}
In the rest of our discussion in this section, we will always work with $\hat\alpha$ 
satisfying this bound.

As in Born-Infeld, it would be convenient for us to define a few dimensionless quantities
\begin{equation}
\bar r = {r\over l}, ~\bar \alpha = {\hat \alpha\over {l^2}}, ~\bar q = {q\over l^2}, ~
{\rm and}~\bar m ={m\over l^2}.
\end{equation}
In terms of these quantities the temperature of the black hole $\bar T$ is given by
\begin{equation}
\bar T = {Tl} = {\bar r^2 + 2 \bar r^4 - {\bar q\over {\bar r^2}}\over
{2 \pi \bar r (\bar r^2 + 2 \bar \alpha)}}.
\label{temp}
\end{equation}

Again, it would be instructive to study the structure of the horizon(s) associated with the 
black hole. Firstly notice that, from (\ref{vblack}), we can write a relation between 
mass and horizon radius as 
\begin{equation}
\bar m = \bar r_h^4 + \bar r_h^2 + \bar \alpha + {\bar q^2\over {\bar r_h}^2},
\end{equation}
where $\bar r_h = r_h/l$ and $r_h$ represents the roots of the equation $V(r) = 0$.
The largest root was defined as $r_+$ earlier.
For generic values of $\bar q$ and $\bar \alpha$, at small $\bar r_h$, $\bar m$ 
decreases 
with $\bar r_h$ while for large $\bar r_h$, it increases as $\bar r_h^4$. In 
between, there is
a single minimum. At the minimum, $\bar r_h$ satisfies
\begin{equation}
2\bar r_h^6 + \bar r_h^4 - \bar q^2 = 0.
\label{extr}
\end{equation}

\begin{figure}[t]
\begin{center}
\begin{psfrags}
\psfrag{a}[][]{$\bar r_h$}
\psfrag{b}[][]{$\bar m$}
\epsfig{file=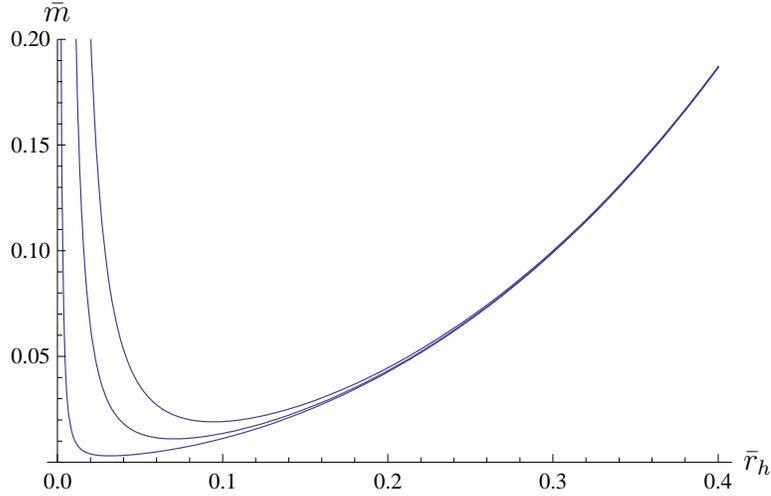, width= 10cm,angle=0}
\end{psfrags}
\vspace{ .1 in }
\caption{This is a plot $\bar m$ with $\bar r$ for $\bar \alpha =.001$ for $\bar q = 
.001, .005, .009$.
As $\bar q$ increases, the critical value of $\bar m$, where the two horizons meet, 
increases.}
\label{m-r}
\end{center}
\end{figure}

The behaviour of $\bar m$ as a function of $\bar r_h$ is shown in Figure (\ref{m-r}). 
It is clear from the 
figure that in general for a given $\bar q$, as long as $\bar m$ is more than a critical 
value, there are two horizons. They will be called inner and outer horizons subsequently.
At the critical value of $\bar m$, when (\ref{extr}) is satisfied, two horizons coincide.
From (\ref{temp}), we see that the temperature of the hole is then zero. We call this
hole as an extremal black hole. When we further decrease the mass, there are no  
horizons. 
One then has a naked singularity.

Let us now consider a three-brane with tension $\Lambda_{\rm br}$ moving in this 
background. Choosing the brane metric as in equation (\ref{frw}), we get the 
Hubble equation of the form \cite{Maeda:2007cb}
\begin{equation}
{\bar \Lambda_{\rm br}\over 3} = \Big( {V(\bar a)\over{\bar a^2}} + \bar H^2 
\Big)\Big[1 + 
{2 
\bar\alpha\over 
3}\Big( {3 - V(\bar a)\over {\bar a^2}} + 2 \bar H^2\Big)\Big]^2.
\label{hub}
\end{equation}
In the above, we have used $\bar a = {a/l}$. $\bar H$, the Hubble parameter,  is given 
by $\bar H = d \bar a/(\bar ad \bar \tau)$ where $\bar \tau = \tau/l$. Furthermore, we
have defined $\bar \Lambda_{\rm br} = l^2 \Lambda_{\rm br}$.

In the limit $\bar \alpha \rightarrow 0$, (\ref{hub}) simplifies and goes over
to the form given in equation (\ref{smal}). The solution for the scale factor for
$\bar \alpha \rightarrow 0$ is then given by (\ref{betainfty}), resulting 
a cyclic universe. Thus, for a vanishing 
$\bar \alpha$, we have non-singular cosmological solutions on the brane. 

Though for finite $\bar\alpha$, it is hard to find an exact closed form solution 
of (\ref{hub}), however, it is easy to see that the bouncing solutions still 
continue to survive. Note that the effective AdS scale for non-zero $\bar \alpha$ 
and large $r$ is given by $L$ as in (\ref{efads}). As before, and for the sake 
of comparison, we make effective brane tension to zero at large $r$ by setting
\begin{equation}
\Lambda_{\rm br} = 3 L^{-2}
\end{equation}
Hence the equation (\ref{hub}) reduces to
\begin{equation}
\Big[{1\over{2 \bar \alpha}} - {1\over{2 \bar\alpha}} \Big(1 - 4 \bar\alpha 
\Big)^{1\over 2}\Big]
 = \Big( {V(\bar a)\over{\bar a^2}} + \bar H^2
\Big)\Big[1 +
{2
\bar\alpha\over
3}\Big( {3 - V(\bar a)\over {\bar a^2}} + 2 \bar H^2\Big)\Big]^2.
\label{hubtwo}
\end{equation}
This is a cubic equation of $\bar H^2$ and can be re-written as
\begin{eqnarray}
{16 \bar \alpha^2\over 9}\bar H^6 &+& {8 \bar \alpha \over 3}\Big(1 + {2 \bar \alpha\over {\bar 
a^2}}\Big) \bar 
H^4
+ {1\over {3 \bar a^4}}\Big( 3 \bar a^2 + 6\bar \alpha - 2 \bar \alpha V\Big)\Big(\bar a^2 + 2 \bar\alpha 
+ 2 \bar \alpha V\Big) \bar H^2 \nonumber\\
&+& {1\over{18 \bar a^6 \bar\alpha}}\Big(9 \bar a^6 ({\sqrt{1 - 4\bar\alpha}} -1) 
+ 18 \bar a^4 \bar\alpha V \nonumber\\
&-& 24 \bar \alpha^2 \bar a^2 (V - 3)V + 8\bar\alpha^3 (V - 3)^2 V\Big) = 0.
\label{hsol}
\end{eqnarray}
This equation can, in principle, be explicitly solved to get $\bar H^2$. The physical domain, however,  corresponds to
the region $\bar H^2 \ge 0$. The points where the equality holds represent the turning points.
Unlike the case of $\bar \alpha = 0$, we have not been able to integrate the above equation 
to get the scale factor $\bar a$ as a function of time in a closed form. However, 
as in the Born-Infeld case, it is possible to
re-write (\ref{hsol}) as 
\begin{equation}
\Big({d \bar a\over{d \bar \tau}}\Big)^2 + U(\bar a) =0,
\label{effc}
\end{equation}
with $U(\bar a)$, the effective potential.
By studying $U(\bar a)$ for different values of parameters, we would be able to infer the nature
of the brane universe. This is what we do in the following.

\begin{figure}[t]
\begin{center}
\begin{psfrags}
\psfrag{x}[][]{$\bar a$}
\psfrag{y}[][]{$U(\bar a)$}
\epsfig{file=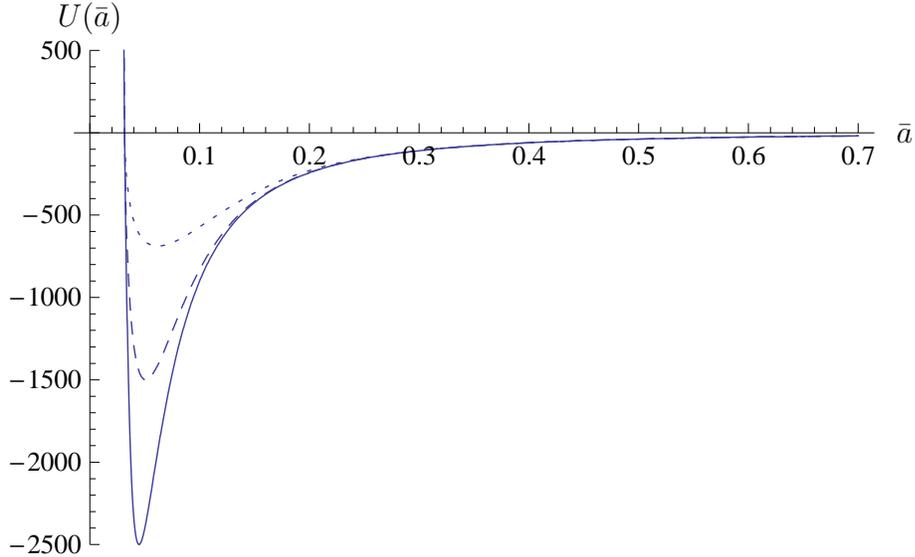, width= 12cm,angle=0}
\end{psfrags}
\vspace{ .1 in }
\caption{A plot of the potential $U(\bar a)$ with $\bar a$. Three curves correspond to 
$\bar \alpha = 10^{-7}$(dotted), $\bar \alpha = 10^{-6}$ (dashed) and the other is for $\bar \alpha =0$.
We have also set $\bar m = 10$ and $\bar q = .1$.
From (\ref{effc}), 
allowed physical region
is $U(\bar a) \le 0$. Since $U(\bar a) = 0$ occurs only at non-zero finite values of $\bar a$, we 
get non-singular cyclic brane universe. }
\label{V-a}
\end{center}
\end{figure}

Figure (\ref{V-a}) shows a plot of the 
effective potential $U(\bar a)$ for different values of $\bar \alpha$ including 
the 
Reissner-Nordstrom black 
hole ($\bar \alpha = 0$). Notice that the potential does not allow the brane to 
hit the 
black hole singularity. It bounces before reaching the singularity and thus 
generates a bouncing 
universe. It can be checked that for large $\bar a$, $U(\bar a)$ changes sign at 
some value of $\bar a$,
becomes positive and asymptotically reaches a constant. We can therefore 
conclude 
that the brane again bounces before reaching an infinite size. This is 
typical of a bouncing cyclic 
universe. However, we must mention here that, as in 
the case of $\bar \alpha = 0$ \cite{Mukherji:2002ft}, 
it can however be checked that for a bounce at small 
$\bar a$ 
to occur, the brane has to cross the inner horizon. 
As in $\bar \alpha = 0$, this may
induce instabilities in the bulk causing an appearance of a 
singularity on the inner horizon \cite{Hovdebo:2003ug}.
For that, one needs to study perturbations on the bulk due to the brane. We leave 
this for a future study.

To conclude, in this note, we have shown the existence of bouncing cosmological 
solutions by studying the dynamics of three-brane in certain black hole 
backgrounds. These black holes are the classical solutions of Born-Infeld
and Gauss-Bonnet theories in the presence of negative cosmological constant.
The analysis guides us to  infer that the charge parameter of the bulk black holes
 play crucial role, inducing bouncing cosmology on the brane. 
More precisely, the charge induces a negative energy density on the brane thereby
preventing the brane to fall in the black hole singularity, and in turn, succeeding to give
singularity-free cosmology on the four-dimensional world. 
Finally, it is important to study the stability of these solutions. We plan to
return to this issue in the future.

\section*{Acknowledgments}

SP thanks Institute of Physics, Bhubaneswar for warm hospitality
on a visit during which the project was substantiated.

\end{document}